\documentstyle[times,epsfig,graphics,amssymb]{mn2e} 

\newcommand{\be}{\begin{equation}} 
\newcommand{\e}{\end{equation}} 
\newcommand{\bear}{\begin{eqnarray}} 
\newcommand{\ear}{\end{eqnarray}} 
 
\newcommand{\f}{\frac}

\def\be{\begin{equation}} 
\def\ee{\end{equation}}

\def\HI{\hbox{H$\scriptstyle\rm I\ $}}

\def\gsim{\lower.5ex\hbox{\gtsima}} 
\def\lsim{\lower.5ex\hbox{\ltsima}} 
\def\gtsima{$\; \buildrel > \over \sim \;$} 
\def\ltsima{$\; \buildrel < \over \sim \;$} 
\def\prosima{$\; \buildrel \propto \over \sim \;$} 
\def\gsim{\lower.5ex\hbox{\gtsima}} 
\def\lsim{\lower.5ex\hbox{\ltsima}} 
\def\simgt{\lower.5ex\hbox{\gtsima}} 
\def\simlt{\lower.5ex\hbox{\ltsima}} 
\def\simpr{\lower.5ex\hbox{\prosima}}

\title[Cosmic density field reconstruction from Ly$\alpha$ forest data] 
{Cosmic density field reconstruction from Ly$\alpha$ forest data}

\author[S. Gallerani, F.~S. Kitaura, A. Ferrara]{ 
S. Gallerani$^{1}$, F.~S. Kitaura$^{2,3}$, A. Ferrara$^{2}$ \\ 
$^1$ INAF-Osservatorio Astronomico di Roma, via di Frascati 33, 00040 Monte Porzio Catone, Italy\\ 
$^2$ Scuola Normale Superiore, Piazza dei Cavalieri 7, 56126, Pisa, Italy \\ 
$^3$ Ludwig-Maximilians Universitat Munchen, Scheinerstr. 1, D-81679, Munich, Germany\\ 
} 
\date{\today} 
\pagerange{\pageref{firstpage}--\pageref{lastpage}} \pubyear{2011} 
\begin{document} 
\maketitle 
\label{firstpage} 
\begin{abstract} 
We present a novel, fast method to recover the density field through the statistics of the transmitted flux in high redshift quasar absorption spectra. The proposed technique requires the computation of the probability distribution function of the transmitted flux ($P_{\rm F}$) in the Ly$\alpha$ forest region and, as a sole assumption, the knowledge of the probability distribution function of the matter density field ($P_{\Delta}$). We show that the probability density conservation of the flux and matter density unveils a flux-density ($F-\Delta$) relation which can be used to invert the Ly$\alpha$ forest without any assumption on the physical properties of the intergalactic medium. We test our inversion method at $z=3$ through the following steps: [i] simulation of a sample of synthetic spectra for which $P_{\Delta}$ is known; [ii] computation of $P_{\rm F}$; [iii] inversion of the Ly$\alpha$ forest through the $F-\Delta$ relation. Our technique, when applied to only 10 observed spectra characterized by a signal-to noise ratio S~/~N~$\geq 100$ provides an exquisite (relative error $\epsilon_{\Delta}\lesssim 12$\% in $\gtrsim 50$\% of the pixels) reconstruction of the density field in $\gtrsim 90$\% of the line of sight. We finally discuss strengths and limitations of the method. 
\end{abstract} 
\section{Introduction} 
The ultraviolet radiation emitted by a quasar can suffer resonant Ly$\alpha$ scattering as it propagates 
through the intergalactic neutral hydrogen. In this process, photons are removed from the line of 
sight resulting in an attenuation of the source flux, the so-called Gunn-Peterson effect (Gunn \& Peterson 1965). At $z\sim 3$, the neutral hydrogen number density is sufficiently small to allow to resolve the wide series of absorption lines, which give rise to the so-called Ly$\alpha$ forest in quasar absorption spectra. Initially, the absorption lines observed in quasar spectra were believed to be produced by an intergalactic population of pressure-confined clouds, photoionized by the integrated quasar flux (Sargent et al. 1980). Nowaday absorbers of the Ly$\alpha$ forest are generally associated to large-scale neutral hydrogen density fluctuations in the warm photoionized IGM. \\ 
This scenario, firstly tested through analytical calculations (Bond et al. 1988; Bi et al. 1992), has received an increased consensus thanks to the detection of a clustering signal both along individual (Cristiani et al. 1995; Lu et al. 1996; Cristiani et al. 1997; Kim et al. 2001) and close pairs of quasar lines of sight, resulting in an estimated size for the absorbers which varies between few hundreds of kpc (Smette et al. 1992; D'Odorico et al. 1998; Rauch et al. 2001; Becker et al. 2004) to few Mpc (Rollinde et al. 2003; Coppolani et al. 2006; D'Odorico et al. 2006). Hydrodynamical cosmological simulations lend further support to this picture, having established the existence of a tight connection between the \HI density field and the dark matter distribution, at least on scales larger than the Jeans lenght of the IGM (Cen et al. 1994; Petitjean, et al. 1995; Zhang et al. 1995; Hernquist et al. 1996; Miralda-Escude' et al. 1996). As a consequence of this result, the Ly$\alpha$ forest has been proposed as a powerful method to study the clustering properties of the dark matter density field and to measure its power spectrum (Croft et al. 1998; Nusser \& Haehnelt 1999; Pichon et al. 2001; Rollinde et al. 2001; Zhan et al. 2003; Viel et al. 2004; McDonald et al. 2005; Zaroubi et al. 2006; Saitta et al. 2008; Bird et al. 2010).\\ 
In this paper, we focus our attention on the reconstruction of the density field along the line of sight (LOS), which represents the starting point for the dark matter power spectrum measurement. We present a novel, fast method to recover the density field along the LOS towards $z\sim 3$ quasar which can be considered alternative and/or complementary to the existing techniques in the literature mentioned above which will be further discussed in this work. 
\section{Modelling the forest}\label{modfor} 
\begin{figure*} 
\centerline{ 
\psfig{figure=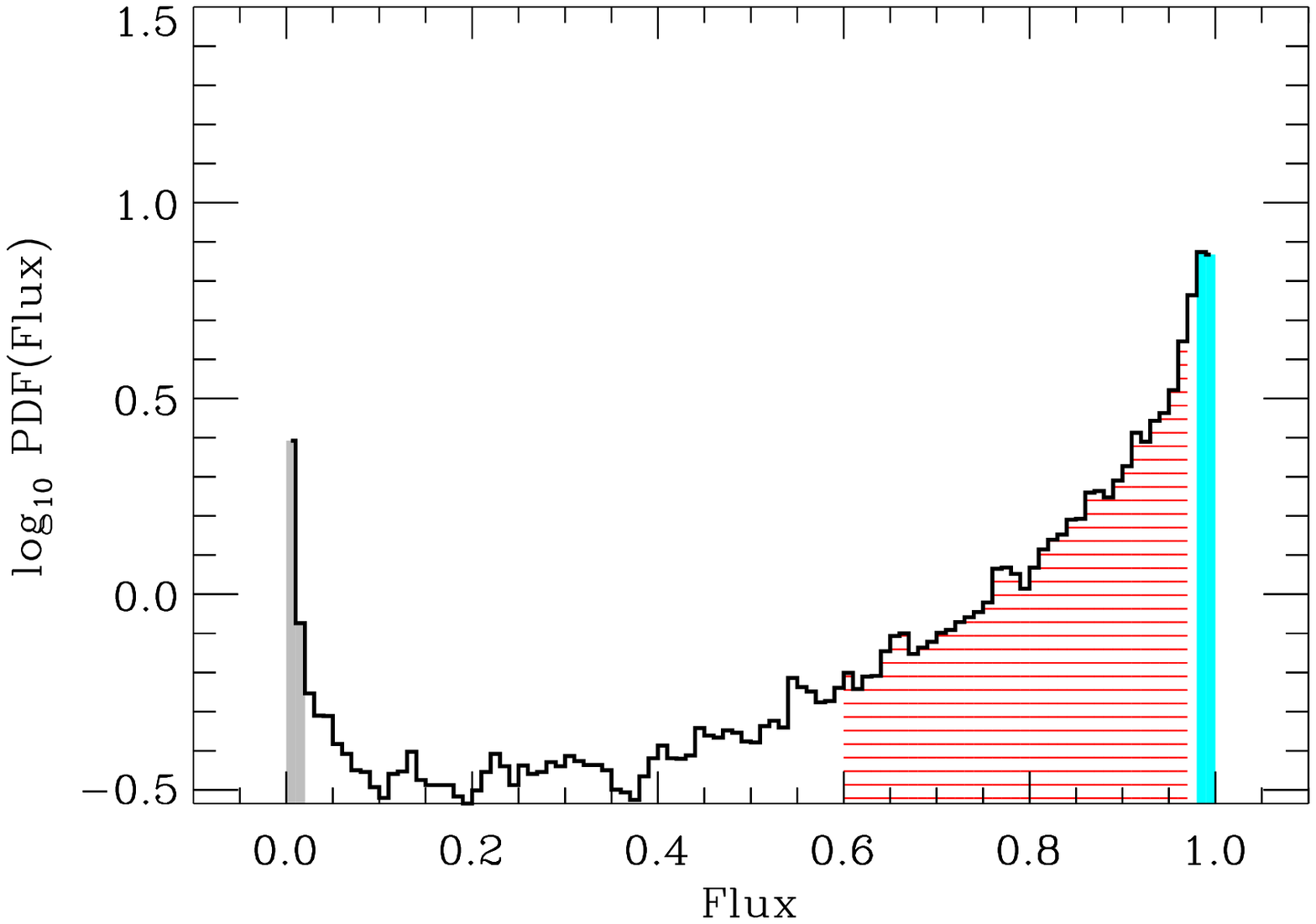,width=7.cm,angle=0} 
\psfig{figure=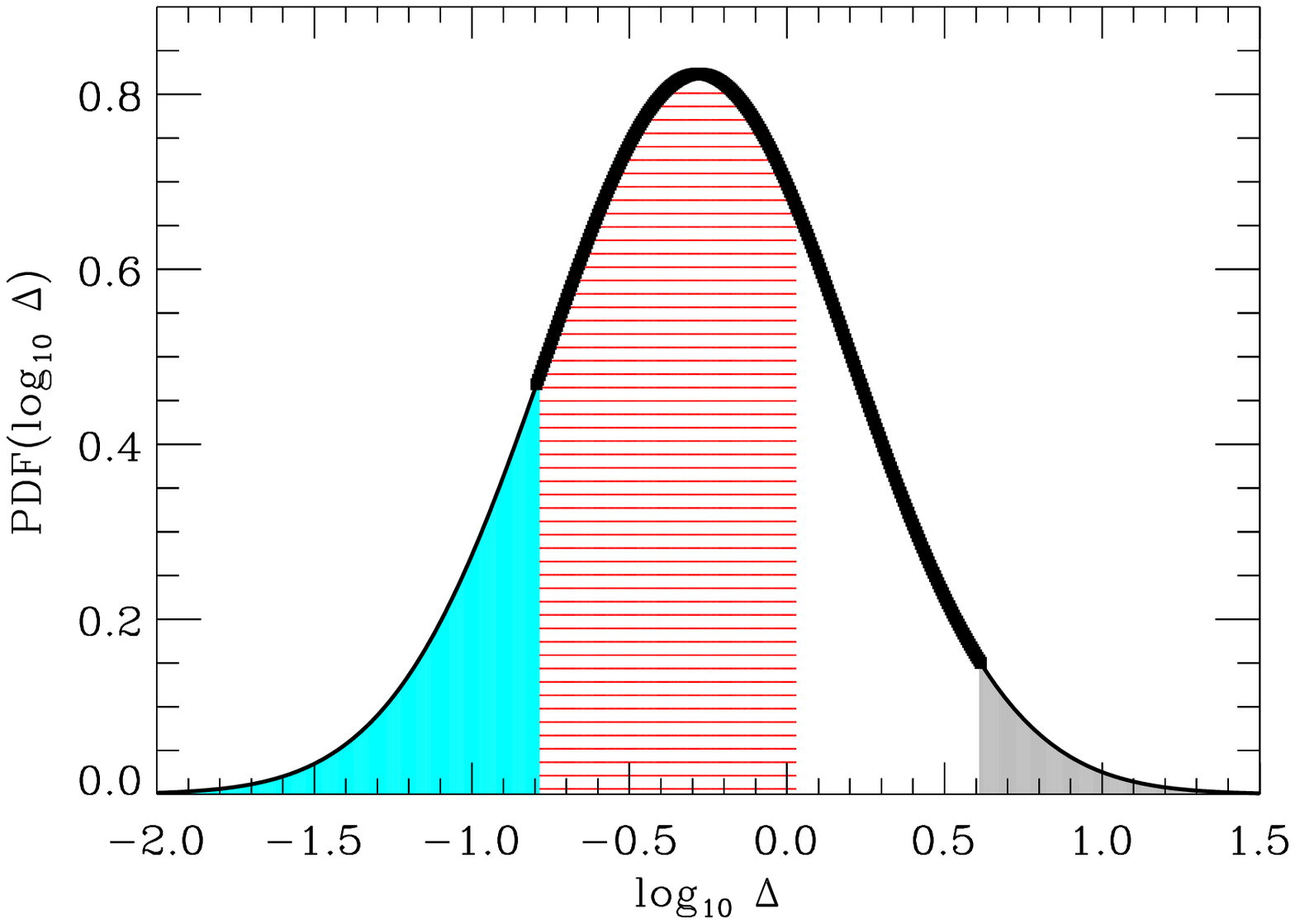,width=7.cm,angle=0} 
} 
\caption{{\it Left panel}: Probability distribution function of the transmitted flux obtained from 10 synthetic quasar absorption spectra at $z=3$. {\it Right panel}:  Probability distribution function of the density field at $z=3$, as predicted by the lognormal model. The thick line shows $P_{\Delta}$ of the recovered density field. The probability of finding a pixel characterized by a transmitted flux above a given $F_{\rm max}$ (cyan shaded region on the left-hand panel) is equal to the probability of finding an overdensity below $\Delta_{\rm b}$ (cyan shaded region on the right-hand panel). Analogously, the probability of finding a pixel characterized by a transmitted flux below a given $F_{\rm min}$ (gray shaded region on the left-hand panel) is equal to the probability of finding an overdensity above $\Delta_{\rm d}$ (gray shaded region on the right-hand panel). In particular, at each pixel characterized by a transmitted flux $F_*$ it can be associated an overdensity $\Delta_*$ such that eq.~(\ref{deltastar1}) is satisfied. In this case, the probability of finding a pixel characterized by a transmitted flux $F_{\rm min}<F_*<F_{\rm max}$ (red hatched region on the left-hand panel) is equal to the probability of finding an overdensity $\Delta_{\rm b}<\Delta_*<\Delta_{\rm d}$ (red hatched region on the right-hand panel).}     
\label{method} 
\end{figure*} 
Let us discretize the LOS towards a distant quasar in a number of pixels $N_{\rm pix}$.  
The transmitted flux due to Ly$\alpha$ absorption in the IGM at a given pixel $i$ is computed from the usual relation  
\begin{equation} 
F(i)=e^{-\tau(i)}\,, 
\end{equation}  
where $\tau(i)$ is the  Ly$\alpha$ absorption optical depth: 
\begin{equation}\label{taueq} 
\tau(i)=c I_{\alpha} \f{\Delta x}{1+z(i)} 
\sum^{N_{\rm pix}}_{j=1} n_{\rm HI}(j) \Phi_{\alpha}[v_{\rm H}(i)-v_{\rm H}(j)]\,, 
\label{lyalphatau} 
\end{equation} 
where $c$ is the light speed, $I_{\alpha}$ is the Ly$\alpha$ cross section, $\Delta x$ is the comoving pixel size, $z(i)$ is the redshift of pixel $i$, $\Phi_{\alpha}$ is the Voigt profile for the Ly$\alpha$ transition, $v_{\rm H}$ is the Hubble velocity\footnote{The velocity of a Ly$\alpha$ forest absorber is generally given by the sum of the Hubble and the peculiar velocities, i.e. $v(i)=v_{\rm H}(i)+v_{\rm pec}(i)$. In this work, we neglect peculiar velocities and we discuss this issue in the Discussion section.}, and $n_{\rm HI}$ is the neutral hydrogen fraction.\\ 
The computation of $n_{\rm HI}$ generally assumes that the low density gas which gives rise to the Ly$\alpha$ forest is approximately in local equilibrium between photoionization and recombination. By neglecting the presence of helium, and assuming that the IGM is highly ionized at the redshift of interest by a uniform ultraviolet background, the photoionization equilibrium condition provides us with the following relation: 
\begin{equation}\label{photoeq} 
n_{\rm HI}(j)=\frac{\alpha[T(j)]}{\Gamma}\{n_{\rm 0}[1+z(j)]^3\Delta(j)\}^2\propto \Delta(j)^{\beta}\,, 
\end{equation}  
with $1.5<\beta<2.0$ (Hui \& Gnedin 1997). In eq. (\ref{photoeq}), $\Gamma$ is the photoionization rate of neutral hydrogen at a given redshift $z(j)$, $\Delta(j)$ is the overdensity at the pixel $j$, $n_{\rm 0}$ is the mean baryon number density at $z=0$, and $\alpha[T(j)]\propto T(j)^{-0.7}$ is the temperature-dependent radiative recombination rate. 
For quasi-linear IGM, where non-linear effects like shock-heating can be neglected, the temperature can be related to the baryonic density through a power-law relation (Hui \& Gnedin 1997): 
\begin{equation}\label{eos} 
T(j)=T_{\rm 0}\Delta(j)^{\gamma-1}\,, 
\end{equation} 
where $T_{\rm 0}$ is the IGM temperature at the mean density at a given redshift $z(j)$ which depends on the reionization history of the Universe, such as the slope $\gamma$ of the equation of state, and $\Gamma$.\\ Besides the $n_{\rm HI}$ computation, the thermal state of the IGM affects the profile of the absorption lines. In fact, the Voigt profile which enters in eq. (\ref{taueq}) is the convolution of a Gaussian and a Lorentzian profile, since it takes into account the thermal and the natural broadening of the absorption line, respectively. For low column density regions, as the ones corresponding to Ly$\alpha$ forest, thermal broadening dominates, and the Voigt function reduces to a simple Gaussian  
\begin{equation} 
\Phi_{\alpha}[v_{\rm H}(i)-v_{\rm H}(j)]=\frac{1}{\sqrt{\pi}~b(j)}  
\exp\left\{-\left[\f{v_{\rm H}(i)-v_{\rm H}(j)}{b(j)}\right]^2\right\}\,, 
\end{equation} 
whose width is set by the temperature-dependent Doppler parameter $b(j)=\sqrt{2 k_{\rm B} T(j)/m_{\rm p}}$, 
with $k_{\rm B}$ and $m_{\rm p}$ being the Boltzmann constant and the proton mass, respectively.\\ 
From the system of equations (1)-(5) the complex relation between the transmitted flux $F$ in quasar absorption spectra and the underlying density field $\Delta$ becomes evident. The determination of the $F-\Delta$ relation not only depends on the parameters describing the physical state of the IGM ($\Gamma; T_{\rm 0}; \gamma$), i.e. on the cosmic reionization history, but also on the convolution of the resulting $n_{\rm HI}$ with the profile of the absorption lines, which in  turn is linked to the thermal state of the gas through the Doppler parameter $b$. In the next Section, we describe how the $F-\Delta$ relation can be efficiently inferred from a statistical analysis of the transmitted flux in quasar absorption spectra.  
\begin{figure*}\label{fdid} 
\centerline{ 
\psfig{figure=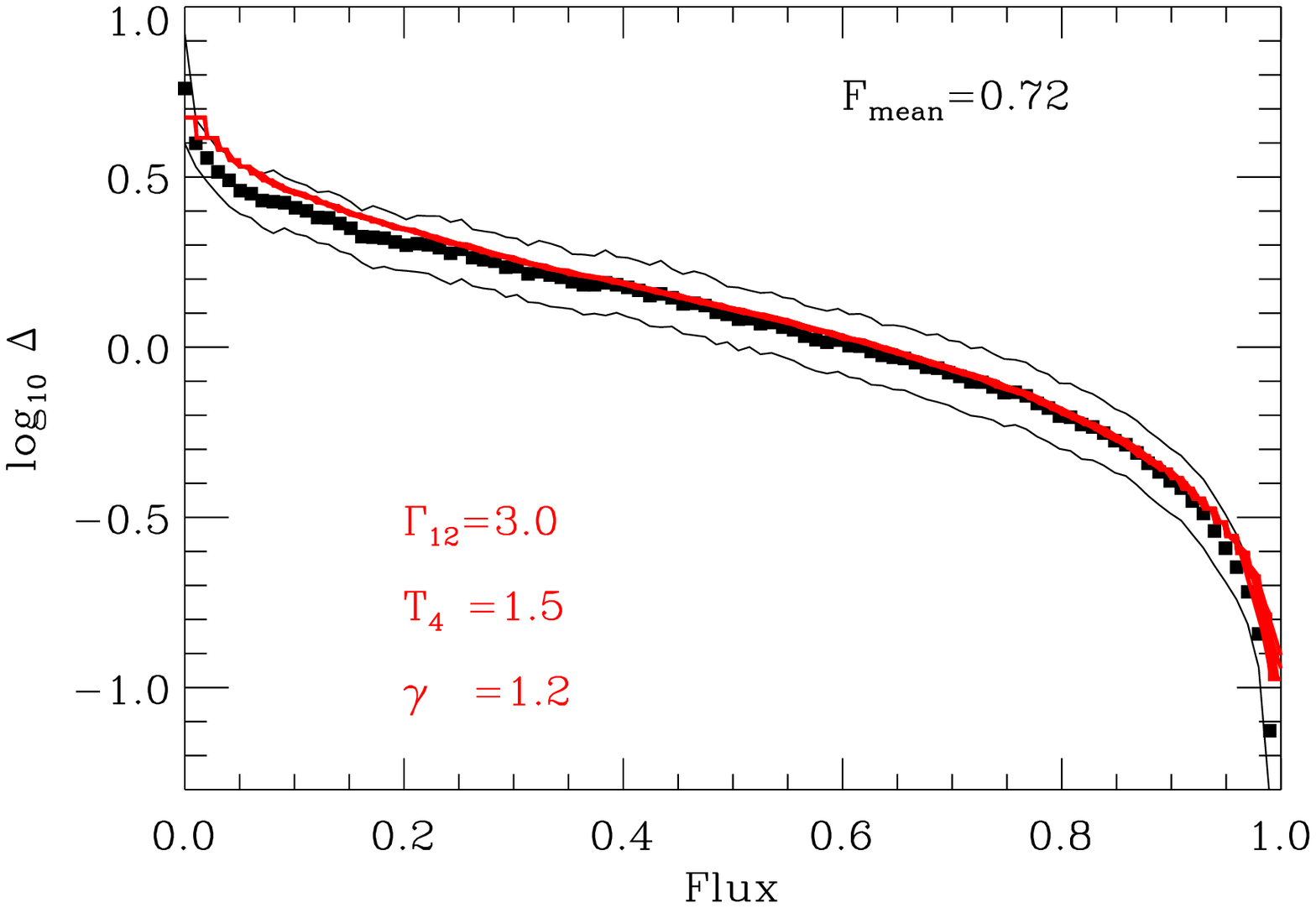,width=6.cm,angle=0} 
$\!\!\!\!$
\psfig{figure=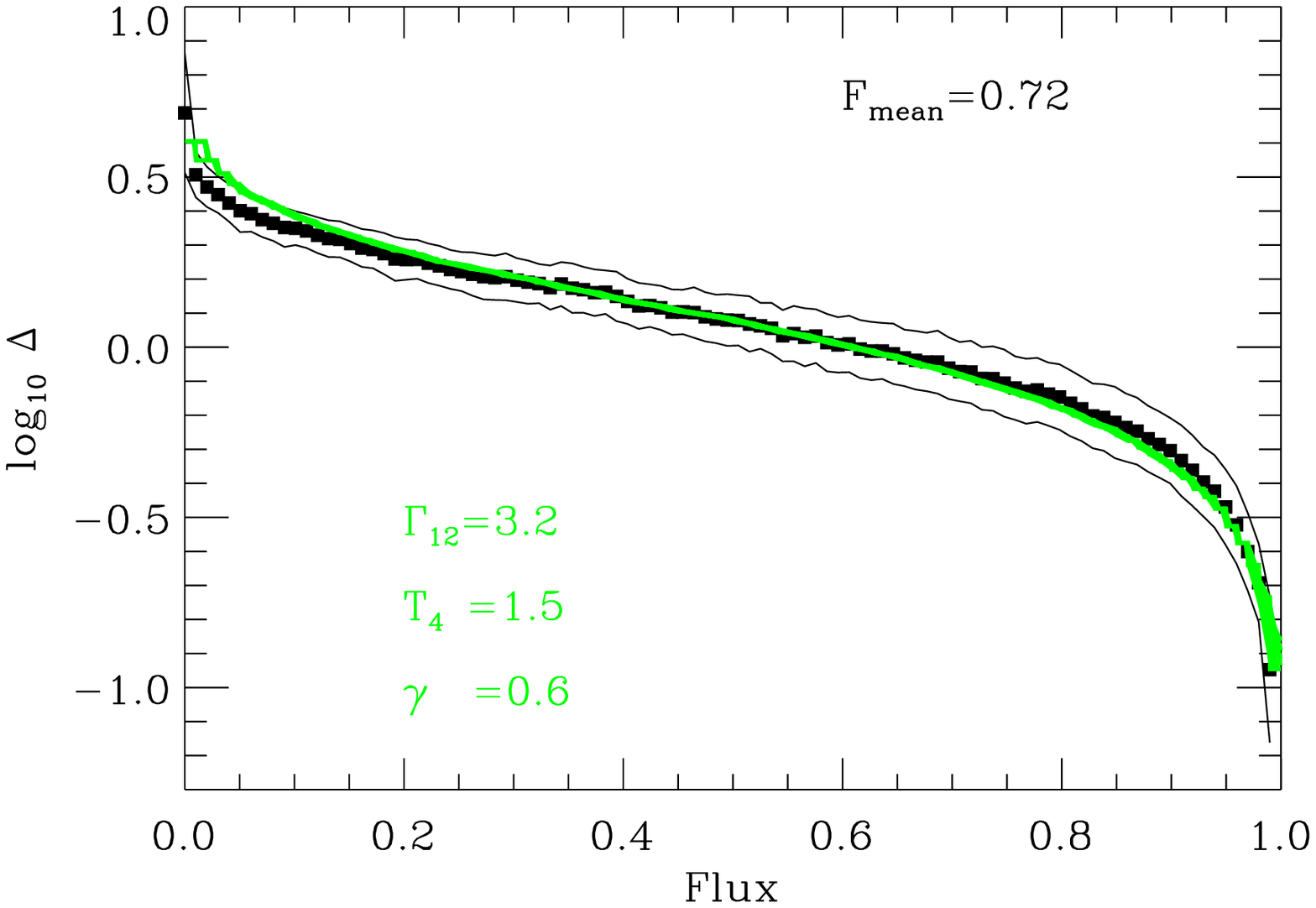,width=6.cm,angle=0} 
$\!\!\!\!\!\!$
\psfig{figure=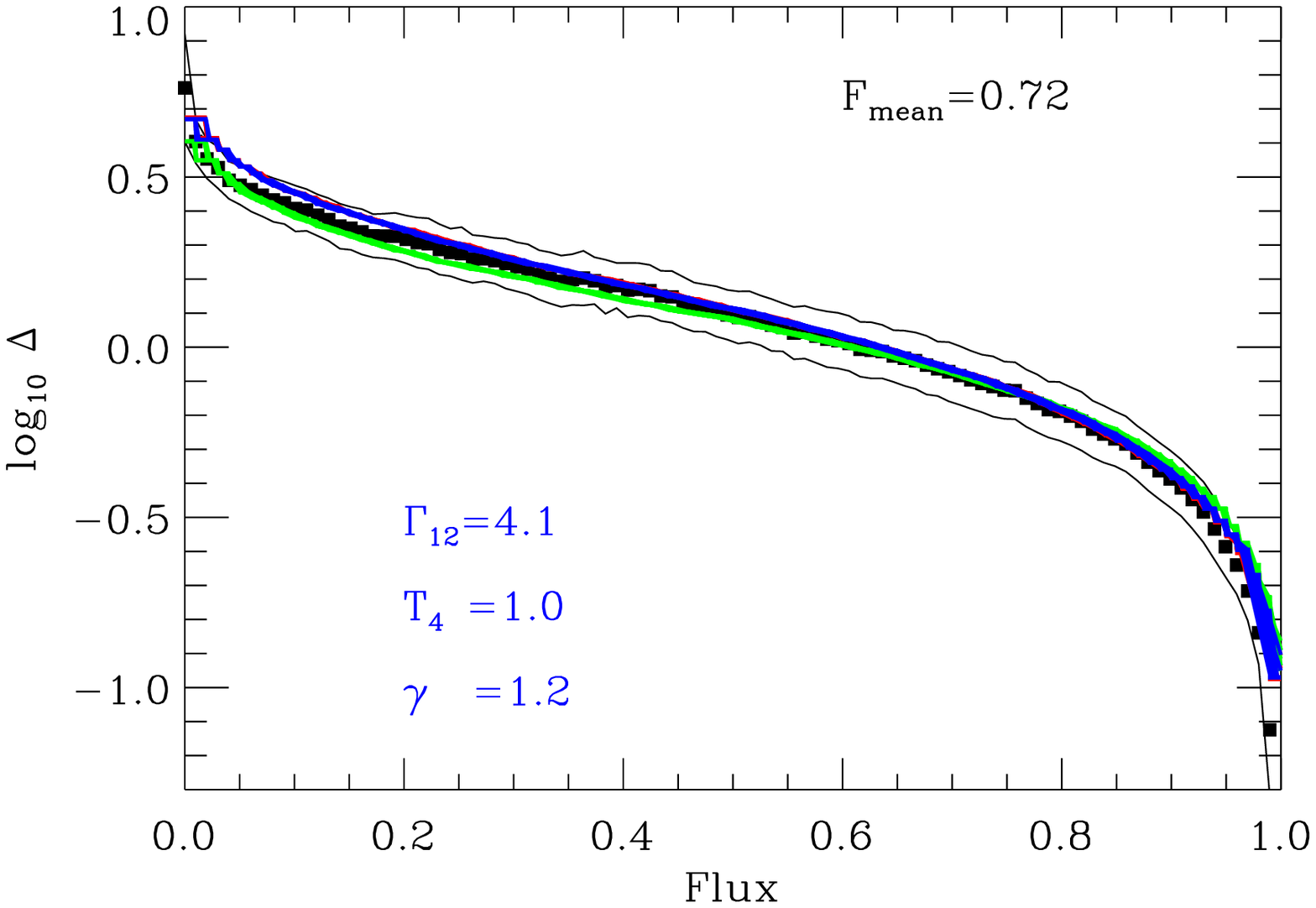,width=6.cm,angle=0} 
} 
\caption{Flux-density relation resulting from flux/density probability conservation is shown for different parameters defining the IGM physical properties, as labelled in the figures. The black squares represent results obtained from 10 synthetic spectra, while coloured lines denote the flux-density relations resulting by solving eq. (\ref{deltastar1}). In the rightmost panel, the flux-density relation is plotted for all the set of parameters considered, which provide the same mean flux.} 
\end{figure*}  
 
\section{Flux-Density relation}\label{secfd} 
Let us assume that the signal-to-noise ratio characterizing an observed quasar absorption spectrum is such that the maximum flux (minimum optical depth) which can be distinguished from full transmission is given by $F_{\rm max}$ ($\tau_{\rm min}$), and that the minimum detectable flux (maximum optical depth) is given by $F_{\rm min}$ ($\tau_{\rm max}$). 
By combining eq. (\ref{taueq})-(\ref{photoeq})-(\ref{eos}), it results that  
$\tau=\tau(z,\beta, \Gamma, T_{\rm 0},\gamma,\Delta)$. 
This means that, at a given redshift $z$, for any $\beta$, and for any given IGM thermal and ionization history, two characteristic overdensities do exist: [i] $\Delta_{\rm b}$, where the subscript ``{\rm b}'' means ``bright'', such that each Ly$\alpha$ absorber sitting on $\Delta<\Delta_{\rm b}$ would provide $\tau(\Delta_{\rm b})\lesssim \tau_{\rm min}$, i.e. full transmission of the quasar continuum ($F>F_{\rm max}$); [ii] $\Delta_{\rm d}$, where the subscript ``{\rm d}'' means ``dark'', such that each overdensity $\Delta>\Delta_{\rm d}$ would provide $\tau(\Delta_{\rm d})\gtrsim \tau_{\rm max}$, therefore completely depleting the flux emitted by the source at the location of the absorber ($F<F_{\rm min}$). \\ 
One of the statistics adopted to analyze quasar spectra is the probability distribution function of the transmitted flux ($P_{\rm F}$), which quantify the number of pixels characterized by a transmitted flux between $F$ and $F+dF$. 
Let us assume that at a given overdensity $\Delta_*$ it is associated a unique value of the transmitted flux $F_*$. In this case, $\Delta_{\rm b}$ and $\Delta_{\rm d}$ can be defined through the probability of finding a pixel characterized by a transmitted flux  $F > F_{\rm max}$, and  $F < F_{\rm min}$, respectively. Once $P_{\rm F}$ is computed from observed spectra, $\Delta_{\rm b}$ and $\Delta_{\rm d}$ can be determined from the following equations:    
\begin{equation}\label{deltabright} 
\int_{F_{\rm max}}^1 P_F~dF=\int_0^{\Delta_{\rm b}}P_{\Delta}~d\Delta\,, 
\end{equation}  
\begin{equation}\label{deltadark} 
\int_0^{F_{\rm min}} P_F~dF=\int_{\Delta_{\rm d}}^{+\infty}P_{\Delta}~d\Delta\,, 
\end{equation}  
where $P_{\Delta}$ is the probability distribution function of the density field. Analogously to eq.~(\ref{deltabright}) and eq.~(\ref{deltadark}), to each pixel characterized by a transmitted flux $F_{\rm min}<F_*<F_{\rm max}$ an overdensity $\Delta_{\rm b}<\Delta_*<\Delta_{\rm d}$ can be associated such that:  
\begin{equation}\label{deltastar1} 
\int_{F_*}^{F_{\rm max}} P_F~dF=\int_{\Delta_{\rm b}}^{\Delta_*}P_{\Delta}~d\Delta\,, 
\end{equation}  
or, equivalently, 
\begin{equation}\label{deltastar2} 
\int_{F_{\rm min}}^{F_*} P_F~dF=\int_{\Delta_*}^{\Delta_{\rm d}} P_{\Delta}~d\Delta\,. 
\end{equation}  
Note that eq.~(\ref{deltabright})-(\ref{deltadark})-(\ref{deltastar1})-(\ref{deltastar2}) are simply obtained by imposing the conservation of the flux/density probability densities. The inference of $\Delta_{\rm b}$, $\Delta_{\rm d}$, $\Delta_*$ only depends on the assumed $P_{\Delta}$ and not on any assumption concerning thermal and ionization histories which are however encoded into the observed $P_{\rm F}$, as we show in the next Section. By solving eq. (\ref{deltastar1}), for each observed value of $F_*$ in terms of $\Delta_*$, the flux-density relation of the Ly$\alpha$ forest can be readily derived. Such $F-\Delta$ relation can be used to invert the Ly$\alpha$ forest, therefore allowing to reconstruct the density field along the LOS.\\   
\section{Inversion procedure}\label{invsec} 
\begin{figure} 
\centerline{ 
\psfig{figure=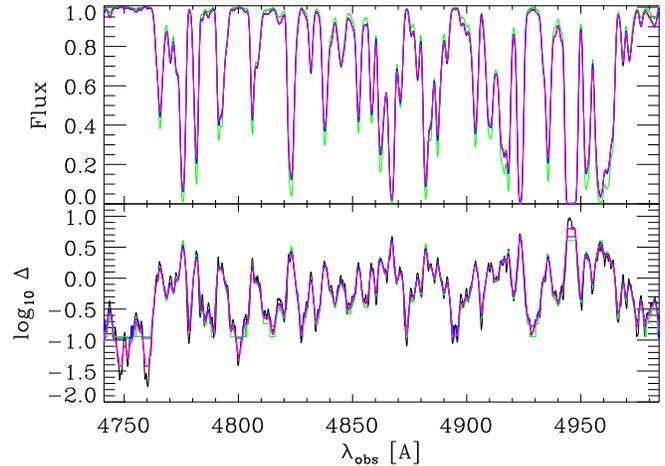,width=9.cm,angle=0} 
} 
\caption{{\it Top panel}: Example of synthetic quasar absorption spectra at $z=3$ as obtained by adopt the following set of parameters: [i] ($\Gamma, T_{\rm 0}, \gamma)=(3.0, 1.5, 1.2$) (red line); [ii] $(3.2, 1.5, 0.6)$ (green line); [iii] $(4.1, 1.0, 1.2)$ (blue line), where $\Gamma$ is reported in units of $10^{-12}~{\rm s}^{-1}$, and $T_{\rm 0}$ in units of $10^4~K$. {\it Bottom panel}: The dotted black line represents the distribution of the overdensities which produce the transmitted flux shown in the top panel. The red line (green/blue) show the recovered density field, as obtained by inverting the flux density relation corresponding to the set of parameters [i] [ii/iii]. The magenta line show the case of a high quality (S/N=1000) spectrum.}
\label{firstrec} 
\end{figure}

We simulate quasar absorption spectra at $z=3$ through the model described in the previous section. In particular, for what concerns the baryonic density distribution along the LOS entering in eq. (\ref{photoeq}), we adopt the method described by Gallerani et al. (2006), whose main features are summarized as follows.  The spatial distribution of the baryonic density field and its correlation with the peculiar velocity field are taken into account adopting the formalism introduced by Bi \& Davidsen (1997). To enter the mildly non-linear regime which characterizes the Ly$\alpha$ forest absorbers we use a lognormal model introduced by Coles \& Jones (1991).\\ 
The thermal and ionization state of the IGM are fixed in such a way that the mean transmitted flux simulated in a sample of lines of sight is $\bar{F}\sim 0.72$, as obtained from real data (Songaila 2004). We adopt the following set of parameters: [i] ($\Gamma, T_{\rm 0}, \gamma)=(3.0, 1.5, 1.2$); [ii] $(3.2, 1.5, 0.6)$; [iii] $(4.1, 1.0, 1.2)$, where $\Gamma$ is reported in units of $10^{-12}~{\rm s}^{-1}$, and $T_{\rm 0}$ in units of $10^4~K$.\\
We add observational artifacts to the synthetic spectra: first, we smooth each simulated spectrum to a resolution R=36000, comparable to the one of high-resolution spectrographs (e.g. HIRES); then we add noise to get a signal-to-noise ratio S/N=100; finally we rebin the noisy synthetic spectra to R=10000, in order to remove small flux fluctuations. Starting from a sample of 10 LOS, we compute the resulting $P_{\rm F}$ for the cases [i], [ii], [iii]. In Fig.~\ref{method}, $P_{\rm F}$ for the case [i] is shown in the left panel, through a solid line.\\ 
We assume the lognormal $P_{\Delta}$ which has been adopted for the synthetic spectra simulations, and which is described by a Gaussian with mean~$=-0.28$, and sigma~$=0.48$ (Fig.~\ref{method}, right panel, solid line). The choice of a lognormal $P_{\Delta}$ is only made for testing purposes, but when dealing with real data any other, perhaps more realistic, prescription for $P_{\Delta}$ can be implemented. For example, Miralda-Escude' et al. 2000 suggested a different functional form, whose agreement with data has not yet fully established (see discussion in Becker et al. 2006).\\ 
Starting from the computed $P_{\rm F}$ and the assumed $P_{\Delta}$, we solve eq.~(\ref{deltastar1}), to establish the flux-density relation for each of the set of parameters considered. The resulting $F-\Delta$ relations\footnote{In this work we adopt the cosmological parameters provided by Komatsu et al. (2010). Uncertainties on the cosmological parameters have negligible effect on the flux-density relation.} are shown through coloured lines in Fig. \ref{fdid} for the case [i] (left-most panel, red line), [ii] (middle panel, green line), [iii] (right-most panel, blue line). In this figure, the filled squares and the solid black lines represent the mean flux and the relative 1$\sigma$ error, respectively, corresponding to a given overdensity, as obtained from our simulations. It can be seen that the resulting $F-\Delta$ relations obtained by solving eq.~(\ref{deltastar1}), agree very well with the relation between the mean flux and the overdensities predicted by the simulations. However, our method does not manage to take into account the dispersion of the $F-\Delta$ relations.\\The reason for this is due to our assumption of a one-to-one correspondence between the transmitted flux observed at a given pixel $i$ and the underlying matter density field at the corresponding redshift $z(i)$. Such an assumption would be valid if the profile of the absorption lines would be described by a Dirac Delta function (i.e. in the ideal case of an infinitely small natural broadening of the line), and in the case of a cold IGM (i.e. in the ideal case of no thermal broadening of the line). In this ideal case the cross section of the Ly$\alpha$ transition is different from zero only in correspondence of the pixel $i$ for which we compute the optical depth. In the real case, to the optical depth of the pixel $i$ do contribute also the adjacent pixels. In particular, the higher is the IGM temperature (i.e. the larger is the doppler parameter), the larger is the number of adjacent pixels responsible for the optical depth of the pixel $i$. Therefore, an observed value of the transmitted flux $F_*$ at the pixel $i$ can be produced by an isolated overdensity $\Delta>\Delta_*$, as well as by a clustering of overdensities $\Delta<\Delta_*$. Our method does not take into account the broadening of the absorption lines, therefore it can not provide the dispersion in the $F-\Delta$ relations; in the discussion section, we will analyze the effect of this limit of our method on the density field reconstruction.\\ 
In Fig.~\ref{firstrec}, we show an example of the simulated density field for a random LOS (bottom panel, black line) and the corresponding transmitted fluxes (top panel) obtained for for the case [i] (red line), [ii] (green line), [iii] (blue line). In the bottom panel of Fig.~\ref{firstrec} the originally adopted density field (black line) is compared with the recovered ones, as obtained by applying the $F-\Delta$ relation to the transmitted flux computed from the set of parameters [i] (red line), [ii] (green line), and [iii] (blue line). This figure shows that the method works quite well\footnote{By computing the relative error $\epsilon_{\Delta}$ of the recovered density field along $\geq 10$ lines of sight, we find that $\gtrsim 50$\% of the pixels are characterized by $\epsilon_{\Delta}\lesssim 0.12$.} for all the set of parameters considered, i.e. almost independently from the physical properties of the IGM. 
This comes out by the fact that the $F-\Delta$ relations obtained by solving eq. (\ref{deltastar1}) in terms of $\Delta_*$ (coloured lines in Fig. \ref{fdid}) are not completely independent on the reionization history, i.e. different set of parameters affects differently the $P_{\rm F}$, therefore providing slightly different $F-\Delta$ relations. This is the reason why by inverting the Ly$\alpha$ forest we can associate the same density field to different values of the transmitted flux, obtaining a reconstruction method which is free of any assumption on the thermal and ionization state of the IGM. In the next section we discuss the strengths and the limits of our method in further details. 
\section{Discussion} 
From the bottom panel of Fig.~\ref{firstrec}, it is evident that the method is overall very successful. Nevertheless some 
minor discrepancie between the original and reconstructed density field remain. In particular, the reconstructed density field is (1) smoother than the original one, and (2) it overestimates (underestimates) the original one in the most underdense (overdense) regions.\\ The accuracy loss (1) arises from the adopted assumption of a one-to-one correspondence between the transmitted flux and the matter density. As already discussed in the previous section, this assumption does not allow us to take into account the clustering of the density field locally, or equivalently the thermal broadening of the absorption lines. This results in a reconstructed density field which is smoother than the original one.\\ 
As for limit (2), this is due to the fact that we are considering both a maximum flux ($F_{\rm max}$) which can be distinguished from full transmission, and a minimum detectable flux ($F_{\rm min}$). As explained in Sec. \ref{secfd}, these maximum and minimum transmitted fluxes translate into a minimum ($\Delta_{\rm b}$) and maximum ($\Delta_{\rm d}$) overdensities which can be recovered. Stated differently, in correspondence of the most underdense regions ($\Delta<\Delta_{\rm b}$) we can only put an upper limit on the recovered density field, i.e. ($\Delta=\Delta_{\rm b}$); analogously, for the most overdense regions ($\Delta>\Delta_{\rm d}$) only a lower limit ocan be set, i.e. ($\Delta=\Delta_{\rm d}$). For a fixed mean transmitted flux, the maximum and minimum recoverable overdensities depend on IGM properties, and are more sensitive\footnote{For an inverted equation of state [iii], $\Delta_{\rm b}$ is higher and $\Delta_{\rm d}$ is lower than in the case of a ``regular'' one. In fact, as can be seen from Fig. \ref{firstrec}, in the case [iii] (green line), the most underdense regions ($\Delta<\Delta_{\rm b}$) tend to be more transmitting ($F>F_{\rm max}$) and the most opaque regions ($\Delta>\Delta_{\rm d}$) to be even more absorbed ($F<F_{\rm min}$)} to changes in $\gamma$ than to $T_0$: [i/ii] ($\Delta_{\rm b};\Delta_{\rm d})=(0.09; 4.90$); [iii] ($\Delta_{\rm b};\Delta_{\rm d})=(0.11; 3.90$). Typical uncertainties on $F_{{\rm mean}}$ at $z=3$ ($<10$\%, Songaila 2004) implies small variations on $\Delta_{{\rm b}}$ and $\Delta_{{\rm d}}$. For example, in the case [i] a mean transmitted flux 10\% higher [lower] than 0.72 results in ($\Delta_{\rm b};\Delta_{\rm d})=(0.11; 6.10$) [($\Delta_{\rm b};\Delta_{\rm d})=(0.08; 3.82$)].\\ 
The existence of $\Delta_{\rm b}$ and $\Delta_{\rm d}$ allows us to recover the density field along $\gtrsim 92$\% of the simulated lines of sight. This result can be further improved by applying our method to very high quality spectra. Up to now we have assumed a $S/N=100$, which means that we are sensitive at 1$\sigma$ level to flux variations up to the second decimal digit. Therefore, we have considered $F_{\rm max}=0.99$ and $F_{\rm min}=0.01$. However, if we consider higher quality data (e.g. S/N=1000), the original density field is better recovered, since the completeness of the signal rises to $\gtrsim 98\%$, being ($\Delta_{\rm b};\Delta_{\rm d})=(0.03; 6.60$). In this case, we obtain a reconstructed signal characterized by a relative error $\epsilon_{\Delta}\lesssim 12$\% in 51\% of the pixels, $12\%<\epsilon_{\Delta}<42\%$ in 39\% of the pixels, and $42\%<\epsilon_{\Delta}<100\%$ in 10\% of the pixels.\\ 
We finally compare our method with previously developed techniques. One of the approaches generally adopted in the density field reconstruction is to define a {\it physical} model relating the observed flux to the underlying density field and then invert this relation. As shown in Sec. \ref{modfor}, any model which allows to predict the transmitted flux resulting from a given density field distribution, at a given redshift $z$, depends on several uncertain parameters ($\beta$, $\Gamma$, $T_0$ $\gamma$). This approach has been pioneered by Nusser \& Haehnelt (1999), hereafter NH99. These authors have developed a technique based on a model of the Ly$\alpha$ forest, which allows to reconstruct the density field along quasar sightlines through an iterative method. Such technique allows to correctly take into account the thermal broadening of the absorption lines, assuming that both $T_0$ and $\gamma$, which affect the Doppler parameter, are known.\\ The NH99 inversion method has been adopted by several authors to study the properties of the IGM and to measure the matter power spectrum from the Ly$\alpha$ forest (e.g. Pichon et al. 2001; Rollinde et al. 2001; Zaroubi et al. 2006). However, it has been recognized that uncertainties on $\beta$, $\gamma$, and $T_0$ result in spurious biases in the recovered density field (NH99; Pichon et al. 2001; Rollinde et al. 2001). The strong dependence of the matter power spectrum on the IGM properties, as inferred from Ly$\alpha$ forest data, has been also confirmed by Bird et al. (2010). Saitta et al. (2008) have proposed another possible approach to reconstruct the density field through the transmitted flux in quasar absorption spectra, named FLO (i.e. From Lines to Overdensities). Also in this case the parameters defining the IGM equation of state need to be assumed. Moreover, while the FLO method seems promising for the reconstruction of the most overdense regions (up to $\Delta\sim 30$), underdensities ($\Delta <1$) tend to be underestimated.\\
Summarizing, the methods available so far in the literature tried to invert the $F-\Delta$ relation for which a large number of assumptions have to be made on the physical state of the IGM. We have shown that the $F-\Delta$ relation can be efficiently inferred from a {\it statistical} analysis of the transmitted flux in quasar absorption spectra drastically reducing the number of assumptions. As far as this work is concerned, we have not considered peculiar velocity effects which influence the quality of the recovery (NH99). However, we note that the same method described in NH99 for the reconstruction in real space can be applied to our recovered density field iteratively. We leave this to a future work. We finally note that so far we have been referring to quasar absorption spectra. However, the method proposed can be applied to GRB absorption spectra as well.
\section{Conclusions} 
A novel, fast method to recover the density field through the statistics of the transmitted flux in high redshift quasar absorption spectra has been introduced. The proposed technique requires the computation of the probability distribution function of the transmitted flux ($P_{\rm F}$) in the Ly$\alpha$ forest region and, as a sole assumption, the knowledge of the probability distribution function of the matter density field ($P_{\Delta}$). We have shown that the probability density conservation of the flux and matter density unveils a flux-density ($F-\Delta$) relation which can be used to invert the Ly$\alpha$ forest, without any assumption on the properties of the IGM.\\ Our inversion method has been then tested at $z=3$ through a semi-analytical model of the Ly$\alpha$ forest which adopts a lognormal $P_{\Delta}$. First of all we have simulated a sample of synthetic spectra varying the properties of the IGM ($\Gamma$; $T_0$; $\gamma$) in such a way that the resulting mean transmitted flux matches observations. Then, we have computed $P_{\rm F}$ for each parameter set considered. Different IGM properties affect differently the $P_{\rm F}$, hence resulting in slightly different $F-\Delta$ relations. This provides a reconstruction method which does not require {\it any assumption} on the thermal and ionization state of the IGM. The proposed method is particularly suitable for the extraction of large scale matter density fields signals from Ly$\alpha$ data, which represents the starting point for the detection of baryonic acoustic oscillations through quasar absortion spectra (McDonald \& Eisenstein 2007; Slosar et al. 2009; Kitaura et al. 2010). In fact, this kind of studies are based on the density field reconstruction on scales smaller than 0.1 Mpc along a large number of lines of sight more than 10 Mpc long. Such requirements are beyond the capability of current simulations, while can be satisfied through our fast, though approximate, technique.
\section*{Acknowledgments}
SG acknowledges a ELTE Budapest Postdoctoral Fellowship, and a SNS Pisa Visiting Scientist Fellowship which have partially supported this work. The authors thank the Intra-European Marie-Curie fellowship with project number 221783 and acronym MCMCLYMAN for supporting this project.
 
\end{document}